\documentstyle[preprint,aps]{revtex}
\tighten
\begin{document}
\draft
\title{Levinson's Theorem for Dirac Particles}
\author{J. Piekarewicz}
\address{Supercomputer Computations Research Institute, \\
         Florida State University, Tallahassee, FL 32306}
\date{\today}
\maketitle

\begin{abstract}
Levinson's theorem for Dirac particles constraints the sum of
the phase shifts at threshold by the total number
of bound states of the Dirac equation. Recently, a stronger
version of Levinson's theorem has been proven in which the value
of the positive- and negative-energy phase shifts are separately
constrained by the number of bound states of an appropriate set
of Schr\"odinger-like equations. In this work we elaborate on
these ideas and show that the stronger form of Levinson's theorem
relates the individual phase shifts directly to the number of
bound states of the Dirac equation having an even or odd number
of nodes. We use a mean-field approximation to Walecka's
scalar-vector model to illustrate this stronger form of Levinson's
theorem. We show that the assignment of bound states to a particular
phase shift should be done, not on the basis of the sign of
the bound-state energy, but rather, in terms of the nodal
structure (even/odd number of nodes) of the bound state.
\end{abstract}

\narrowtext

\section{Introduction}
\label{secintro}

	One of the most beautiful results in scattering theory
was proven by Levinson in 1949~\cite{levin49}. Levinson's theorem
relates the number of bound states in a specific angular-momentum
channel ($N_{l}$) to the value of the phase shift ($\delta_{l}$)
at threshold
\begin{equation}
  \delta_{l}(0)=N_{l}\pi \;.
 \label{nrellevin}
\end{equation}
In arriving at this form of Levinson's theorem the modulo-$\pi$
ambiguity in the definition of the phase shift has been resolved by
demanding that the phase shift approaches zero (and not only a multiple
of $\pi$) in the high-energy limit. For simplicity, we have assumed,
and will assume hereafter, the absence of zero-energy resonances.

	Finding a generalization of Levinson's theorem to Dirac
particles, however, has proven to be a subtle task. Initially, it
was believed that the generalization of Levinson's theorem related
the difference of positive- and negative-energy bound states to the
corresponding difference of positive- and negative-energy phase shifts
at threshold~\cite{ni79}. It was also suggested that the modulo-$\pi$
ambiguity in the definition of the phase shift could be resolved, as
in the nonrelativistic case, by allowing the phase shift to vanish in
the high-energy ($E \rightarrow \pm \infty$) limit. These results,
however, were later found to be incorrect and it was only until 1985
that a correct generalization of Levinson's theorem to Dirac particles
was found by Ma and Ni~\cite{mani85}.
They showed that the correct form of Levinson's theorem for Dirac
particles relates the sum (and not the difference) of the phase
shifts at threshold to the total number of bound states~\cite{mani85}
\begin{equation}
  \delta_{\kappa}(E\!=\!+M)+\delta_{\kappa}(E\!=\!-M)=N_{\kappa}\pi \;.
 \label{rellevin}
\end{equation}
In the above equation $\kappa$ is the generalized angular momentum
channel ($j=|\kappa|-1/2$) and $N_{\kappa}$ is the total number
of (positive- plus negative-energy) bound states in that channel.
In this form of Levinson's theorem one implicitly assumes that the
modulo-$\pi$ ambiguity in the definition of the phase shift
has already been resolved. However, in contrast to the nonrelativistic
case, where one is free to define the high-energy limit of the phase
shift to be zero~\cite{newton66,taylor72}, the asymptotic value for
the phase shift is given by~\cite{parzen50,barthe67},
\begin{equation}
  \lim_{E\rightarrow \pm\infty}\delta_{\kappa}(E)=
  \mp\int_{0}^{\infty} V(r) \, dr \;,
 \label{relasym}
\end{equation}
where $V(r)$ is the timelike component of a Lorentz-vector potential.
In particular, this last relation implies
\begin{equation}
  \delta_{\kappa}(E\!\rightarrow\!+\infty)+
  \delta_{\kappa}(E\!\rightarrow\!-\infty)=0 \;,
 \label{relasymsum}
\end{equation}
and indicates that it is only the sum, but not the individual phase
shifts, that vanishes in the high-energy limit. Yet, Eq.~(\ref{relasym})
establishes that the asymptotic behavior of the of the individual
(positive- and negative-energy) phase shifts is known. In contrast,
the behavior of the individual phase shifts at threshold is (as opposed
to their sum) not constrained by Levinson's theorem. Recently, however,
Poliatzky was able to prove a stronger version of Levinson's theorem
by showing that the individual phase shifts at threshold are
related to the number of bound states of an appropriate set of
Schr\"odinger-like equations which coincide with the Dirac equation
in the limit of zero momentum~\cite{polia92}. In contrast to previous
expectations~\cite{barthe67}, however, the phase shift at threshold
with a particular sign of the energy does not coincide (in general)
with the number of bound states of the Dirac equation having the same
sign of the energy, i.e.,
\begin{equation}
  \delta_{\kappa}(E=\pm M) \ne N_{\kappa}^{(\pm)}\pi \;.
\end{equation}
As we will show later, this ``naive'' expectation is upset
whenever the potential becomes sufficiently strong to cause the
binding energy of certain states to exceed the rest mass of the
particle.

	In the present work we show, using some of the ideas
developed by Poliatzky~\cite{polia92}, that the behavior of
the phase shift at threshold with a particular sign of the
energy is directly related to the number of bound states of
the Dirac equation having a specific nodal structure. We
will establish that the assignment of bound states to a
particular phase shift should be done, not on the
basis of the sign of the bound-state energy but, instead,
by determining if the combined number of nodes in the upper and
lower components of the bound-state wavefunction is even or odd.
As we shall see, this criterion is inextricably linked to the
orthogonality and boundary conditions satisfied by the eigenstates
of the Dirac equation.

Our paper has been organized as follows. In Sec.~\ref{secformal}
we review most of the formalism needed to prove the stronger
form of Levinson's theorem. Particular emphasis is placed on the
boundary conditions satisfied by eigenstates of the Dirac equation
which are ultimately responsible for the nodal structure of the
bound state. In Sec.~\ref{secres} we illustrate Levinson's theorem
using a mean-field approximation to the Walecka model as an example.
In particular, we show that the ``naive'' relation between the phase
shift at threshold and the corresponding number of bound states with
the same sign of the energy is upset whenever the potential becomes
strong enough to generate binding energies that exceed the rest mass
of the particle. Finally, in Sec.~\ref{secconcl} we offer our
conclusions and speculate on future applications of Levinson's theorem.

\section{Formalism}
\label{secformal}

	In the presence of a spherically symmetric potential the
eigenstates of the Dirac equation can be classified according to
a generalized angular momentum $\kappa$ and can be written in a two
component representation
\begin{equation}
  \Psi_{E\kappa m}({\bf r})=
   \left[
    \begin{array}{c}
      \phantom{i}
      \displaystyle{g_{E\kappa}(r) \over r}
      {\cal{Y}}_{+\kappa m}(\hat{\bf{r}})    \\
       i
      \displaystyle{f_{E\kappa}(r) \over r}
      {\cal{Y}}_{-\kappa m}(\hat{\bf{r}})
     \end{array}
   \right] \;,
 \label{dirwf}
\end{equation}
with the upper and lower components expressed in terms of
spin-spherical harmonics
\begin{equation}
  {\cal{Y}}_{\kappa m}(\hat{\bf{r}}) \equiv
  \langle\hat{\bf{r}}|l{\scriptstyle{1\over 2}}jm\rangle  \;, \quad
  j=|\kappa|-{1\over 2} \;, \quad
  l=\cases{  \kappa    & if $\kappa > 0$; \cr
           -(1+\kappa) & if $\kappa < 0$. \cr}
 \label{curlyy}
\end{equation}
For a particle moving in the presence of Lorentz scalar ($S$)
and time-like vector ($V$) potentials the Dirac equation
\begin{equation}
  \Big[
   {\bf \alpha}\cdot{\bf P} + \beta M + \beta S(r) + V(r)
  \Big]\Psi_{E\kappa m}({\bf r}) =
  E\Psi_{E\kappa m}({\bf r}) \;,
 \label{direq}
\end{equation}
reduces to a set of first order, coupled differential equations
\begin{mathletters}
 \label{direqab}
 \begin{eqnarray}
  \left( {d \over dr} + {\kappa \over r} \right) g_{E\kappa}(r) &=&
  \Big[M^{\star}(r) + E^{\star}(r) \Big] f_{E\kappa}(r) \;,
  \label{direqa}  \\
  \left( {d \over dr} - {\kappa \over r} \right) f_{E\kappa}(r) &=&
  \Big[M^{\star}(r) - E^{\star}(r) \Big] g_{E\kappa}(r) \;,
  \label{direqb}
 \end{eqnarray}
\end{mathletters}
written in terms of a local mass and a local energy defined by
\begin{equation}
   M^{\star}(r) \equiv M + S(r) \;, \quad
   E^{\star}(r) \equiv E - V(r) \;.
 \label{mstar}
\end{equation}

	It is a well known fact that the above set of coupled
equations can be reduced to a single second-order differential
equation identical in form to a Schr\"odinger equation. For example,
one can use Eq.~(\ref{direqa}) to express the lower component
of the Dirac wavefunction in terms of the upper component
\begin{mathletters}
 \label{gfplus}
 \begin{eqnarray}
   f_{E\kappa}(r) &=& \left[
                       {1 \over M^{\star}(r) + E^{\star}(r)}
                      \right]
                      \left(
                       {d \over dr} + {\kappa \over r}
                      \right) g_{E\kappa}(r) \;,
   \label{fplus}                                  \\
   g_{E\kappa}(r) &=& \left[\xi(r)\right]^{1/2}u_{E\kappa}(r) =
                      \left[
                      {M^{\star}(r) + E^{\star}(r)  \over M+E}
                      \right]^{1/2} u_{E\kappa}(r) \;,
   \label{gplus}
 \end{eqnarray}
\end{mathletters}
which upon substitution into Eq.~(\ref{direqb}) yields the
following Schr\"odinger-like equation for the (transformed)
upper component
\begin{equation}
  \left[
   {d^{2} \over dr^{2}} + p^{2} -
   {\kappa(\kappa+1) \over r^{2}} -U_{\rm eff}(E,\kappa;r)
  \right] u_{E\kappa}(r) = 0 \;; \quad
  \left(p^{2}\equiv E^{2}-M^{2}\right) \;.
 \label{schreqp}
\end{equation}
The transformation factor $\left[\xi(r)\right]^{1/2}$
relating the Dirac upper component ($g_{E\kappa}$) to the
Schr\"odinger-like component ($u_{E\kappa}$) was introduced
in order to remove all terms linear in the derivative of
$g_{E\kappa}$. The effective potential, which has now acquired
an energy dependence due to the transformation,
\begin{equation}
  U_{\rm eff}(E,\kappa;r) \equiv
    U_{\rm c}(E;r) -
    \kappa U_{\rm so}(E;r) +
    U_{\rm{\scriptscriptstyle D}}(E;r) \;,
 \label{ueff}
\end{equation}
is defined in terms of central, spin-orbit, and Darwin contributions
that are, respectively, given by
\begin{mathletters}
 \begin{eqnarray}
  U_{\rm c}(E;r)  &=& 2M
   \left[
    \left( S(r)+{E\over M}V(r) \right) +
    \left( {S^{2}(r)-V^{2}(r) \over 2M} \right)
   \right]   \;,
   \label{ucent}                \\
  U_{\rm so}(E;r) &=&
    -{1 \over r} \left[{1 \over \xi(r)}{d\xi(r) \over dr}\right] \;,
   \label{uso}                \\
  U_{\rm {\scriptscriptstyle D}}(E;r) &=&
    {3 \over 4}\left[{1 \over \xi(r)}{d\xi(r) \over dr}\right]^{2} -
    {1 \over 2}\left[{1 \over \xi(r)}{d^{2}\xi(r) \over dr^{2}}\right] \;.
   \label{udarwin}
 \end{eqnarray}
\end{mathletters}
In particular, the energy dependence displayed by the (central)
potential now makes evident the high-energy limit of the phase
shifts [Eq.~(\ref{relasym})] through the well known asymptotic
behavior of the Schr\"odinger phase shift~\cite{newton66,taylor72}
\begin{equation}
  \lim_{E\rightarrow \pm\infty}\delta_{\kappa}(E)= -{M \over p}
  \int_{0}^{\infty} V_{\rm eff}(r) \, dr \rightarrow
  \mp\int_{0}^{\infty} V(r) \, dr \;.
 \label{nrelasym}
\end{equation}
Thus, the asymptotic behavior of the phase shift can
ultimately be traced to the Lorentz transformation properties
of the vector potential. In addition, this relation
indicates that the high-energy limit of the phase shift is
unaffected by the presence of a Lorentz-scalar potential.
Moreover, this finding contradicts the nonrelativistic
notion that a distortion-free description of the projectile
should be appropriate at large enough energies.

	In reducing the original couple set of equations to an
effective Schr\"odinger-like equation one could have, alternatively,
chosen to eliminate the upper component of the Dirac wavefunction
in favor of the lower component. This can be achieved, because
of the structure of the equations [see Eq.~(\ref{direqab})],
by simply changing the sign of $\kappa$ and of all timelike
quantities (i.e., $E \rightarrow -E$ and $V \rightarrow -V$).
These modifications give rise to the following relations
\begin{mathletters}
 \label{gfminus}
 \begin{eqnarray}
   g_{E\kappa}(r) &=& \left[
                       {1 \over M^{\star}(r) - E^{\star}(r)}
                      \right]
                      \left(
                       {d \over dr} - {\kappa \over r}
                      \right) f_{E\kappa}(r) \;,
   \label{gminus}                                  \\
   f_{E\kappa}(r) &=& \left[\bar{\xi}(r)\right]^{1/2}v_{E\kappa}(r) =
                      \left[
                      {M^{\star}(r) - E^{\star}(r)  \over M-E}
                      \right]^{1/2} v_{E\kappa}(r) \;,
   \label{fminus}
 \end{eqnarray}
\end{mathletters}
and generate the corresponding Schr\"odinger-like equation for
the (transformed) lower component
\begin{equation}
  \left[
   {d^{2} \over dr^{2}} + p^{2} -
   {\bar{\kappa}(\bar{\kappa}+1) \over r^{2}} -
   \overline{U}_{\rm eff}(E,\bar{\kappa};r)
  \right] v_{E\kappa}(r) = 0 \;.
 \label{schreqm}
\end{equation}
In the above equation we have defined
$\bar{\kappa}\equiv-\kappa$ and introduced the
effective potential $\overline{U}_{\rm eff}$ that is obtained
from $U_{\rm eff}$ by changing the sign of all timelike quantities.

	The system of equations Eqs.~(\ref{gfplus}) and
(\ref{schreqp}), or equivalently Eqs.~(\ref{gfminus}) and
(\ref{schreqm}), with the appropriate boundary conditions,
are equivalent to the original set of coupled equations
[Eq.~(\ref {direqab})]. Unfortunately, the nonrelativistic
form of Levinson's theorem can not be applied to these equations.
The difficulty stems from the fact that the theorem is not valid
once the effective potential has acquired an energy dependence.
Fortunately, Poliatzky has recently been able to show how to
overcome this difficulty~\cite{polia92}. He has proven that for
an equivalent set of Schr\"odinger-like equations,
\begin{mathletters}
 \label{polia}
 \begin{eqnarray}
  \left[
   {d^{2} \over dr^{2}} + p^{2} -
   {\kappa(\kappa+1) \over r^{2}} -U_{\rm eff}(E\!=\!+M,\kappa;r)
  \right] \tilde{u}_{E\kappa}(r) &=& 0 \;,
  \label{poliap}                        \\
  \left[
   {d^{2} \over dr^{2}} + p^{2} -
   {\bar{\kappa}(\bar{\kappa}+1) \over r^{2}} -
   \overline{U}_{\rm eff}(E\!=\!-M,\bar{\kappa};r)
  \right] \tilde{v}_{E\kappa}(r) &=& 0 \;,
  \label{poliam}
 \end{eqnarray}
\end{mathletters}
obtained from replacing the energy-dependence of the effective
potential by its value at threshold, the nonrelativistic form of
Levinson's theorem can indeed be applied~\cite{polia92}. This
stronger form of Levinson's theorem constrains, not only the sum
of the phase shifts [Eq.~(\ref{rellevin}], but, in addition, the
behavior of the individual phase shifts at threshold
\begin{mathletters}
 \label{levinpm}
  \begin{eqnarray}
   \delta_{\kappa}(E\!=\!+M) &=&
   \delta_{l}^{(+)}(p=0)=n_{l}^{(+)}\pi   \;,
   \label{levinp}                         \\
   \delta_{\kappa}(E\!=\!-M) &=&
   \delta_{\bar{l}}^{(-)}(p=0)=n_{\bar{l}}^{(-)}\pi   \;.
   \label{levinm}
  \end{eqnarray}
\end{mathletters}
In the above equations the orbital angular momentum
$l$ and $\bar{l}$ are related, respectively, to
$\kappa$ and $\bar{\kappa}$ through Eq.~(\ref{curlyy}).
In addition, $n_{l}^{(+)}$ and $n_{\bar{l}}^{(-)}$ are
the number of (nonrelativistic) bound states supported by
the energy-independent potentials
($U_{\rm eff}$ and $\overline{U}_{\rm eff}$ respectively).
The equality between phase shifts is a direct consequence
of the equivalence of the Dirac equation to
Eq.~(\ref{poliap}) and Eq.~(\ref{poliam}) at the
corresponding thresholds. The relation to the number of
bound states follows directly from applying the
nonrelativistic form of Levinson's theorem which is
now valid because the potentials have become independent
of energy.

	We now show that the consistency of the above set
of equations enables one to relate the value of the phase
shifts at threshold to the number of bound states of the
Dirac equation having, either, an even or odd number of nodes.
Essential to the proof is the understanding of the asymptotic
behavior (both as $r\!\rightarrow\!0$ and as $r\!\rightarrow\!\infty$)
of the ratio of lower to upper components of the Dirac wavefunction.

For small values of $r$ the upper and lower components of the
Dirac wavefunction can be, respectively, written as
\begin{equation}
  g_{E\kappa}(r) \simeq A\hat{\jmath}_{l}\;(p_{0}^{\star}r) \;; \quad
  f_{E\kappa}(r) \simeq B\hat{\jmath}_{\bar{l}}\;(p_{0}^{\star}r) \;,
  \label{gfsmallr}
\end{equation}
where $\hat{\jmath}_{l}(z)\equiv z{\jmath}_{l}(z)$ is the
Ricatti-Bessel function (i.e., spherical Bessel function
times its argument)~\cite{taylor72}. The coefficients $A, B$
and the wavenumber $p_{0}^{\star}$ are determined (after using
the Dirac equation), by solving the secular
equation
\begin{equation}
 \left|
  \begin{array}{cc}
    p_{0}^{\star}\;{\rm sgn}{\kappa} & -M^{\star}_{0}-E^{\star}_{0} \\
                                     &                              \\
   -M^{\star}_{0}+E^{\star}_{0}      & -p_{0}^{\star}\;{\rm sgn}{\kappa}
  \end{array}
 \right| = 0 \;,
 \label{secular}
\end{equation}
which results in
\begin{equation}
    p_{0}^{\star} = \sqrt{E^{\star2}_{0}-M^{\star2}_{0}} \;,
 \label{pstar}
\end{equation}
\begin{equation}
    \left({B \over A}\right) = +{\rm sgn}{\kappa}
    \left({p_{0}^{\star} \over M^{\star}_{0}+E^{\star}_{0}}\right) \;,
     \quad {\rm or} \quad
    \left({A \over B}\right) = -{\rm sgn}{\kappa}
    \left({p_{0}^{\star} \over M^{\star}_{0}-E^{\star}_{0}}\right) \;,
 \label{abstar}
\end{equation}
where we have defined ${\rm sgn}{\kappa}=\kappa/|\kappa|$,
$M^{\star}_{0}=M^{\star}(r\!=\!0)$, and
$E^{\star}_{0}=E^{\star}(r\!=\!0)$.

The asymptotic ($r \rightarrow \infty$) behavior of the
wavefunction can be obtain in a similar fashion. In this case,
however, one must distinguish between scattering states
($E>+M$ or $E<-M$) and bound states ($-M<E<+M$). For scattering
states the asymptotic behavior is given by
\begin{equation}
  g_{E\kappa}(r) \simeq C
  \sin\Big[pr-{l\pi \over 2}+\delta_{\kappa}(E)\Big] \;; \quad
  f_{E\kappa}(r) \simeq D
  \sin\Big[pr-{\bar{l}\pi \over 2}+\delta_{\kappa}(E)\Big] \;,
  \label{gfsbigr}
\end{equation}
and yields the following relations,
\begin{equation}
    p = \sqrt{E^{2}-M^{2}} \;,
 \label{pfree}
\end{equation}
\begin{equation}
    \left({D \over C}\right) =
     {\rm sgn}{\kappa}\left({p \over E+M}\right) \;,
     \quad {\rm or} \quad
    \left({C \over D}\right) =
     {\rm sgn}{\kappa}\left({p \over E-M}\right) \;.
 \label{cdfree}
\end{equation}
For bound states, on the other hand, the asymptotic behavior is
given by
\begin{equation}
  g_{E\kappa}(r) \simeq C e^{-pr} \;; \quad
  f_{E\kappa}(r) \simeq D e^{-pr} \;,
  \label{gfbigr}
\end{equation}
which in turn implies
\begin{equation}
    p = \sqrt{M^{2}-E^{2}} \;,
 \label{punstar}
\end{equation}
\begin{equation}
    \left({D \over C}\right) =   - \left({p \over M+E}\right)  \;,
     \quad {\rm or} \quad
    \left({C \over D}\right) =   - \left({p \over M-E}\right)  \;.
 \label{cdunstar}
\end{equation}
In particular, this last relation establishes (because $-M<E<+M$)
that the ratio of the two bound-state components is always negative
irrespective of the value of $\kappa$.

Having established the boundary conditions satisfied by the
two components of the Dirac equation we now turn to the set
of equations that have determined the threshold behavior of
the phase shifts. We start by inspecting Eq.~(\ref{gplus})
which defined the relation between $g_{E\kappa}$ and $u_{E\kappa}$
and ultimately determined the value of $\delta_{\kappa}(E\!=\!+M)$.
The fact that both $g_{E\kappa}$ and $u_{E\kappa}$ satisfy real
differential equations with real boundary conditions~\cite{polia92},
demands that the transformation factor
$\xi(r)=\big[M^{\star}(r) + E^{\star}(r)\big]/\big[M+E\big]$
be positive for all values of $r$ and in particular positive
for $r=0$. This implies that bound states of the Dirac equation
satisfying this condition, and therefore being ``counted''
by $\delta_{\kappa}(E\!=\!+M)$, have the sign of the ratio
of the two Dirac components at the origin determined solely
by the sign of $\kappa$, i.e., from Eq.~(\ref{abstar}) we obtain
\begin{eqnarray*}
    {\rm sgn}
     \left[{f_{E\kappa}(r) \over g_{E\kappa}(r)}
     \right]_{r \rightarrow 0} =
    {\rm sgn} \left({B \over A}\right) = {\rm sgn}{\kappa}   \;.
\end{eqnarray*}
Since, in addition, the ratio of the two Dirac components is
always negative at large separations [see Eq.~(\ref{cdunstar})],
these relations constrain the combined number of nodes in the
bound-state wavefunction (i.e., the number of nodes in $g_{E\kappa}$
plus the number of nodes in $f_{E\kappa}$) to be even for
$\kappa \!<\! 0$ and odd for $\kappa \!>\! 0$. In the case of
$\kappa \!>\! 0$, for example, the ratio is positive near the
origin and the lower component must change sign an odd number
of times more than the upper component in order for the ratio
to become negative at large separations.

Relating the behavior of $\delta_{\kappa}(E\!=\!-M)$ to the
number of bound states with the complimentary nodal structure
now follows from Eq.~(\ref{fminus}). Again, the fact that both
$f_{E\kappa}$ and $v_{E\kappa}$ satisfy real differential equations
with real boundary conditions forces the transformation factor
$\bar{\xi}(r)=\big[M^{\star}(r) - E^{\star}(r)\big]/\big[M-E\big]$
to be positive for all values of $r$. Consequently, the behavior
of the ratio of Dirac wavefunctions at the origin is now determined,
not by the sign of $\kappa$ but, instead, by the sign of
$\bar{\kappa}=-\kappa$, i.e.,
\begin{eqnarray*}
    {\rm sgn}
     \left[{g_{E\kappa}(r) \over f_{E\kappa}(r)}
     \right]_{r \rightarrow 0} =
    {\rm sgn} \left({A \over B}\right) = -{\rm sgn}{\kappa}   \;.
\end{eqnarray*}
Consequently, bound states of the Dirac equation satisfying the
above relation, and hence being counted by
$\delta_{\kappa}(E\!=\!-M)$, must
have a combined number of nodes that is now even for $\kappa\!>\!0$
and odd for $\kappa\!<\!0$.
Thus, the boundary conditions have proved essential in
establishing the connection between the threshold behavior
of the phase shifts and the number of bound states of the Dirac
equation with a specific nodal structure.

We can now formulate the stronger version of Levinson's theorem
for a Dirac particle as follows. If the modulo-$\pi$ ambiguity in
the definition of the phase shifts is resolved by demanding the
fulfillment of the high-energy relations
\begin{equation}
  \lim_{E\rightarrow \pm\infty}\delta_{\kappa}(E)=
  \mp\int_{0}^{\infty} V(r) \, dr \;,
\end{equation}
and if $N_{\kappa}^{(e)}(N_{\kappa}^{(o)})$ represents
the number of bound states of the Dirac equation having
an even(odd) number of nodes, then the phase shifts at
threshold are given by
\begin{equation}
  \delta_{\kappa}(E=\pm M)=
    \cases{
      N_{\kappa}^{ (
      {{\lower1.5pt\hbox{$\scriptstyle e$}}
        \atop
       {\raise0.5pt\hbox{$\scriptstyle o$}}} )}\pi &
      if $\kappa < 0$; \cr
      N_{\kappa}^{ (
      {{\lower1.5pt\hbox{$\scriptstyle o$}}
        \atop
       {\raise0.5pt\hbox{$\scriptstyle e$}}} )}\pi &
      if $\kappa > 0$. \cr}
 \label{rlevin}
\end{equation}
So far we have ignored the possible existence of zero-energy
resonances. The presence of zero-energy resonances can
only occur in channels having, either, $\kappa=-1$ or
$\kappa=+1$ and their existence modify the above relations
by the addition of a factor of $\pi/2$ in
the value of
$\delta_{\kappa}(E\!=\!+M)$ or
$\delta_{\kappa}(E\!=\!-M)$
respectively~\cite{ni79,mani85,polia92}.

Deeply rooted in the proof of Levinson's theorem is the
completeness of eigenstates of a Hermitian Hamiltonian.
The completeness property establishes that the appearance of
a bound state must always be accompanied by the corresponding
disappearance of a state in the continuum.
Alternatively, the existence of an attractive potential
does not alter the (infinite) number of states but simply
``pulls down'' some scattering states into the bound-state
region~\cite{ni79,mani85}.
The phase shift at threshold may, thus, be regarded as
a bound-state ``counter'' that triggers every time a
scattering state is pulled down into the bound-state region.

Examined against this background, and with the added constraint
imposed by orthogonality, it may not be surprising that
the phase shifts at the two different thresholds ($E=\pm M$)
are related to the bound states of the Dirac equation with a
complimentary nodal structure. As the potential becomes strong
enough to support the existence of bound states, the phase shifts
at threshold will ``trigger'' every time a scattering state is
pull down into the bound-state region. States moving into the
bound-state region from different ends of the continuum, however,
must necessarily have a different nodal structure. This is a
consequence of the orthogonality of states with different
energies. For example, consider the first state (i.e., the most
deeply bound) that has been pulled down from the negative-energy
continuum and the corresponding one from the positive-energy
continuum. These are the bound states that are expected to have
the fewest number of nodes [see, e.g., Fig.~\ref{figtwo}]. These
two states, however, can not be simultaneously free of nodes;
otherwise they could not be orthogonal and still satisfy the
asymptotic condition
[Eq.~(\ref{cdunstar})]. Therefore, all states that have been pulled
down from the negative-energy continuum must have a complimentary
nodal structure to those states that have originated in the
positive-energy continuum. Furthermore, this nodal structure is
preserved even when the potential is strong enough to cause the
binding energy of some of the states to exceed the rest mass of the
particle. Therefore, it is the nodal structure (even or odd
number of nodes) and not the energy (positive or negative)
that determines the behavior of the phase shift at threshold.

Implicit in the above derivation of the stronger form of Levinson's
theorem is the assumption that the wavenumber in the interior
($p_{0}^{\star}$) is real. This fact conforms to the intuitive
notion that a bound-state wavefunction oscillates in the interior
and falls off exponentially in the exterior. In particular, this
implies that
$\big(M^{\star}_{0}+E^{\star}_{0}\big)$ and
$\big(M^{\star}_{0}-E^{\star}_{0}\big)$ must have opposite signs;
a fact that follows directly from Eq.~(\ref{pstar}), i.e.,
\begin{equation}
  -p^{\star2}_{0}=
  \Big(M^{\star2}_{0}-E^{\star2}_{0}\Big) =
  \Big(M^{\star}_{0}+E^{\star}_{0}\Big)
  \Big(M^{\star}_{0}-E^{\star}_{0}\Big) < 0 \;.
\end{equation}
Consequently, all bound states can be classified according to,
either, $\big(M^{\star}_{0}+E^{\star}_{0}\big)>0$ [those related
to $\delta_{\kappa}(+M)$] or according to
$\big(M^{\star}_{0}-E^{\star}_{0}\big)>0$
[those related to $\delta_{\kappa}(-M)$].

There are, however, some very interesting bound states for
which the interior wavenumber, as defined by Eq.~(\ref{pstar}),
is purely imaginary. These bound states, which appear in the
context of strong-coupling field theories,
display exponential behavior, not only at large distances,
but, in addition, in the (interior) region where one
customarily observes oscillating behavior~\cite{bardeen75,camp76}.
These shell-like states are driven by a strong scalar potential
that generates a strong binding energy and, thus, a rapid
variation of the wavefunction.
The existence of nodes (which leads to an
increase in the kinetic energy), of an angular momentum barrier,
or of an additional vector potential are all factors that may
destroy the shell-like nature of these states~\cite{bardeen75}.
These node-free states are, thus, believed to exist only
in channels with $\kappa=\pm 1$ and, as we show later, do
not seem to violate the stronger form of Levinson's theorem.

We now proceed to illustrate some of the ideas developed so far
by means of some examples.

\section{Results}
\label{secres}
	To illustrate Levinson's theorem we use a mean-field
approximation to the Walecka model~\cite{wal74,serwal86}.
The Walecka model is a relativistic quantum field theory
of nucleons interacting via the exchange of scalar and
vector mesons. The ground-state of the system is obtained from
solving self-consistently a set of mean-field (Hartree)
equations~\cite{horser81}. The outcome of the calculation is a
set of single-particle Dirac orbitals together with scalar and
(timelike) vector mean fields. In Fig.~{\ref{figone}} we show the
self-consistently determined mean fields for ${}^{40}$Ca. The
potentials are strong ($\sim M/2$) and opposite in sign in order
to reproduce the weak-binding energy but strong spin-orbit
splitting characteristic of single-particle nucleon
(i.e., positive-energy) states.

	We have used these spherically-symmetric potentials
to generate all bound states of the Dirac equation having
$\kappa\!=\!-1$. In Fig.~\ref{figtwo} we display the
single-particle spectrum generated from these mean-field
potentials by scaling, both scalar and vector fields, by factors
of 0.6, 1.0 (unscaled) and 1.4 respectively. The two numbers
enclosed in brackets correspond, respectively, to the number
of nodes in the upper ($g_{E\kappa}$) and lower ($f_{E\kappa}$)
components of the bound-state wavefunction. The fact that the vector
field, but not the scalar, changes sign under charge conjugation leads
to spectroscopy of negative-energy states that is driven, in contrast
to the positive-energy states, by a very strong central attraction and
weak spin-orbit splitting. In particular, for $\lambda=1.4$, the state
with the least number of odd nodes (and, thus, ``originally'' in the
negative-energy continuum) has crossed into the positive-energy region.
Figures~\ref{figthree}, \ref{figfour}, and \ref{figfive},
display the energy dependence of the positive- and negative-energy
phase shifts and illustrate the stronger form of Levinson's theorem
(see also Table~\ref{tableone}).
These figures show how the phase shifts at threshold act as
bound-state counters once the modulo-$\pi$ ambiguity in the definition
of the phase shift has been resolved according to Eq.~(\ref{relasym}).
For the first two cases ($\lambda$=0.6, 1.0) the potential (although
quite strong) is not strong enough to bind any Dirac orbital by more
than the rest mass of the nucleon. Hence, the classification of bound
states can be done according to their nodal structure or, equivalently,
according to the sign of the energy, i.e.,
\begin{mathletters}
 \label{naive}
 \begin{eqnarray}
   \delta_{\kappa}(E\!=\!+M) &=& N_{\kappa}^{(e)}=N_{\kappa}^{(+)} \;, \\
   \delta_{\kappa}(E\!=\!-M) &=& N_{\kappa}^{(o)}=N_{\kappa}^{(-)} \;.
 \end{eqnarray}
\end{mathletters}
For $\lambda=1.4$, however, the binding energy of one bound-state
orbital (the one with one single node) exceeds the rest mass of the
nucleon and it becomes inappropriate to count bound states according
to the sign of the energy (see Table~\ref{tableone}), i.e.,
 \begin{equation}
   \delta_{\kappa}(E\!=\!-M) = N_{\kappa}^{(o)}\ne N_{\kappa}^{(-)} \;.
 \end{equation}

	So far we have only studied bound-state orbitals that
display (the traditional) oscillating behavior inside the region
of the potential and exponential falloff in the exterior. We now
turn to the very interesting case of shell-like bound states. For
simplicity, we consider a relativistic Lorentz-scalar square well.
The width $(c)$ of the potential is fixed at $cM=20$. The strength
of the potential, on the other hand, is varied from
$0\!<\!|S_{0}|/M\!<\!2$.
Fig.~\ref{figsix} shows the $\kappa\!=\!-1$ single-particle spectrum
along with the threshold behavior of the positive-energy phase shift
as a function of the scalar strength. Bound states may be obtained
(graphically) by finding the intersection of a circle of radius
$R^{2}=(M^{2}-M^{\star 2})c^{2}$ with tangent-like lines. In
particular, this expression reduces, in the weak-coupling
limit ($|S_{0}|/M\!<<\!1$), to the well-know nonrelativistic relation
\begin{equation}
  R^{2}=\left[M^{2}-M^{\star 2}\right]c^{2} \rightarrow
        2M|S_{0}|\,c^{2} \;.
\end{equation}
Hence the behavior of the spectrum is, at least for $|S_{0}|/M\!<\!1$
easily understood. As the strength of the potential increases
the radius $R$ increases accordingly and gives rise to the formation
of bound states. All these states are characterized by oscillating
behavior inside the region of the well and the appearance of any
bound state is accompanied by a corresponding increase
in the value of the phase shift at threshold. As $|S_{0}|/M=1$
(and $M^{\star}=0$) the radius attains its maximum value and
leads to the maximum number of bound states that can be supported
by this ($cM=20$) potential. As the strength of the potential increases
even further, however, the radius $R$ decreases leading to a weaker
binding and ultimately to the disappearance of all but one bound state.
This interesting dynamics is nicely reflected in the threshold
behavior of the positive-energy phase shift. In fact, since for pure
scalar potentials the binding energy never exceeds the rest mass of
the particle, even the naive expectation for Levinson's
theorem [Eq.~(\ref{naive})] is satisfied.

The sole remaining bound
state is characterized by having a negative effective mass inside
the region of the potential and satisfies $0<E<|M^{\star}|=-M^{\star}$.
In particular, the effective wavenumber $p_{0}^{\star}$, as defined
in Eq.~(\ref{pstar}), is purely imaginary and leads, as shown in
Fig.~\ref{figseven}, to the formation of the shell-like state.
Since the nodal structure of this state remains unchanged
during the oscillating-like to exponential-like transition, the
existence of this shell-like state is still properly accounted for
by the value of the phase shift at threshold.

Notice that for this shell-like state the transformation factor
$\xi(r)$, relating the upper component of the Dirac wavefunction
to the corresponding Schr\"odinger-like wavefunction
[Eq.~(\ref{gplus})], is purely imaginary in the interior
and purely real in the exterior. This suggests that, although it
should be possible to obtain this shell-like solution from solving
the equivalent Schr\"odinger-like equation, one should exercise
care in enforcing these very peculiar set of boundary conditions.

\section{Conclusions}
\label{secconcl}

Levinson's theorem for Dirac particles relates the sum of the
positive- and negative-energy phase shifts at threshold to the
total number of bound states of the Dirac equation. Recently,
Poliatzky was able to prove a stronger version of Levinson's
theorem in which the value of each individual phase shift
at threshold is related to the number of bound states of
a pair of Schr\"odinger-like equations which coincide with the
Dirac equation at zero momentum.
In this paper we have elaborated on some of these ideas and
have shown that the stronger version of Levinson's theorem
relates the value of the Dirac phase shift at threshold
to the number of bound states of the Dirac equation
with a specific nodal structure (i.e., even or odd number of nodes).

We have shown that the simple picture of Levinson's theorem
developed in the nonrelativistic context and based on the
completeness relation is preserved in the relativistic case.
This picture suggests that the existence of an attractive
potential does not alter the number of states but simply
pulls down some scattering states into the bound-state
region. The merit of Levinson's theorem is to identify the
phase shift at threshold as the bound-state counter.

We have illustrated the stronger form of Levinson's theorem
using a mean-field approximation to the Walecka model. We
have explicitly shown that the  ``naive'' nonrelativistic
generalization of Levinson's theorem, namely,
$\delta_{\kappa}(E\!=\!\pm M)=N_{\kappa}^{(\pm)}\pi$, is
violated as soon as the binding energy exceeds the
rest mass of the particle. The nodal structure of the bound
states, however, is robust and remains unchanged even
when the qualitative structure of the bound state is
modified (e.g., shell-like solutions).

One essential feature that must be addressed in the study
of Levinson's theorem for Dirac particles is the
asymptotic behavior of the phase shift [Eq.~(\ref{relasym})].
The fact that the phase shift remains finite
even at very large energies is a relativistic effect that
might lead to interesting consequences.
For example, many coincidence $(e,e'p)$ experiments have
already been proposed at CEBAF in the hope of identifying novel
behavior in the propagation of nucleons through the nuclear medium
(color transparency). To date,
most relativistic analyses of the $(e,e'p)$ reaction employ
an ejectile wavefunction obtained from solving a Dirac equation
having Lorentz scalar and (timelike) vector potentials.
Because most of the analyses conducted so far have been restricted
to low-energy ejectiles, the consequences of the high-energy
behavior of the phase shift have not yet been fully explored.
As the energy of the ejectile increases, as seems to be
required for the onset of color transparency, it will
become essential to understand the relativistic effects
associated with the high-energy propagation of the ejectile.

\acknowledgments
This research was supported by the Florida State University
Supercomputer Computations Research Institute and U.S. Department
of Energy contracts DE-FC05-85ER250000, DE-FG05-92ER40750.

\begin{figure}
 \caption{Self-consistent scalar and vector potentials for
          ${}^{40}$Ca obtained from a mean-field approximation
          to the Walecka model.}
 \label{figone}
\end{figure}
\begin{figure}
 \caption{Single-particle spectrum for the $\kappa\!=\!-1$
          channel in ${}^{40}$Ca for different values of the
          scaling parameter $\lambda$.}
 \label{figtwo}
\end{figure}
\begin{figure}
 \caption{Positive- and negative-energy ($\kappa\!=\!-1$) phase
          shifts as a function of energy for $\lambda=0.6$. The
          dashes represent the asymptotic values of the phase
          shift (i.e., $\pm$ the integral of the vector potential).}
 \label{figthree}
\end{figure}
\begin{figure}
 \caption{Positive- and negative-energy ($\kappa\!=\!-1$) phase
          shifts as a function of energy for $\lambda=1.0$. The
          dashes represent the asymptotic values of the phase
          shift (i.e., $\pm$ the integral of the vector potential).}
 \label{figfour}
\end{figure}
\begin{figure}
 \caption{Positive- and negative-energy ($\kappa\!=\!-1$) phase
          shifts as a function of energy for $\lambda=1.4$. The
          dashes represent the asymptotic values of the phase
          shift (i.e., $\pm$ the integral of the vector potential).}
 \label{figfive}
\end{figure}
\begin{figure}
 \caption{Single-particle spectrum and threshold behavior of
          the positive-energy phase shift as a function of the
          strength of the scalar square-well potential
          ($\kappa\!=\!-1$).}
 \label{figsix}
\end{figure}
\begin{figure}
 \caption{Upper ($g$) and lower ($f$) components of the
          lowest energy $\kappa\!=\!-1$ bound state for
          three different values of the strength of the
          scalar square-well potential.}
 \label{figseven}
\end{figure}

 \mediumtext
 \begin{table}
  \caption{Illustration of Levinson's theorem for $\kappa=-1$
           using a mean-field approximation to the Walecka model
           for ${}^{40}$Ca.}
   \begin{tabular}{ccccccc}
    $\lambda$ &
    $\delta_{\kappa}(+M)/\pi$ & $N_{\kappa}^{(e)}$ & $N_{\kappa}^{(+)}$ &
    $\delta_{\kappa}(-M)/\pi$ & $N_{\kappa}^{(o)}$ & $N_{\kappa}^{(-)}$ \\
        \tableline
    0.20 & 1 & 1 & 1 & 3  & 3  & 3 \\
    0.60 & 2 & 2 & 2 & 6  & 6  & 6 \\
    1.00 & 2 & 2 & 2 & 8  & 8  & 8 \\
    1.40 & 2 & 2 & 3 & 10 & 10 & 9 \\
   \end{tabular}
  \label{tableone}
 \end{table}

\end{document}